%% file: article.tex
\documentclass{PoS}

\usepackage[english]{babel}
\usepackage[utf8]{inputenc}
\usepackage[T1]{fontenc}
\usepackage{amsmath,amssymb,braket,slashed,mathrsfs}

\usepackage[squaren,Gray,cdot,binary]{SIunits}
\usepackage{graphicx}
\graphicspath{{./figures/}}
\usepackage[numbers,square]{natbib}

\newcommand{\cf}{\textit{cf.}~}
\newcommand{\eg}{\textit{e.g.}~}
\newcommand{\ie}{\textit{i.e.}~}
\renewcommand{\bar}{\overline}
\DeclareMathOperator{\bigo}{O}
\newcommand{\fdiff}{\mathrm{D}}
\newcommand{\diff}{\mathrm{d}}
\newcommand{\dm}{\delta m}
\newcommand{\DM}{\Delta M}
\newcommand{\DQCDM}{\Delta_{\QCD}M}
\newcommand{\DQEDM}{\Delta_{\QED}M}
\renewcommand{\epsilon}{\varepsilon}
\newcommand{\QCD}{\mathrm{QCD}}
\newcommand{\QED}{\mathrm{QED}}
\DeclareMathOperator{\SU}{SU}
\DeclareMathOperator{\timeor}{T}
\DeclareMathOperator{\tr}{tr}
\DeclareMathOperator{\U}{U}
\newcommand{\eV}{\electronvolt}
\newcommand{\fm}{\femto\meter}
\renewcommand{\eqref}[1]{(\ref{#1})}
\newcommand{\secref}[1]{section \ref{#1}}
\newcommand{\tabref}[1]{table \ref{#1}}
\newcommand{\figref}[1]{figure \ref{#1}}

\title{Review on the inclusion of isospin breaking effects in lattice 
calculations}
\ShortTitle{Review on the inclusion of isospin breaking effects in lattice 
calculations}
\author{\speaker{Antonin Portelli}\\
        School of Physics \&\ Astronomy, University of Southampton, 
        SO17 1BJ, UK\\
        E-mail: \email{a.portelli@soton.ac.uk}}
\abstract{Isospin symmetry is explicitly broken in the Standard Model by the
non-zero differences of mass and electric charge between the up and down
quarks. Both of these corrections are expected to have a comparable size of the
order of one percent relatively to hadronic energies. Although these
contributions are small, they play a crucial role in hadronic and nuclear
physics. In this review we explain how to properly define QCD and QED on a
finite and discrete space-time so that isospin corrections to hadronic
observables can be computed \emph{ab-initio}. We then consider the different
approaches to compute lattice correlation functions of QCD and QED observables.
Finally we summarise the actual lattice results concerning the isospin
corrections to the light hadron spectrum.}
\FullConference{2013 Kaon Physics International Conference,\\
		29 April-1 May 2013\\
		University of Michigan, Ann Arbor, Michigan - USA}

\begin{document}

\section{Motivations}
In an isospin symmetric world, the up ($u$) and down ($d$) quarks are identical
particles. It is known (\cf\tabref{tab:udquarks}) than in Nature isospin symmetry
is explicitly broken by the non-zero mass and electric charge differences of
the $u$ and $d$ quarks. However, the effects of this breaking are expected to
be small relative to typical strong interaction energies such as hadron masses.
Indeed, from \citep{PDG2012} it is clear that the light quark mass mass
difference $\dm=m_u-m_d$ represents one percent or less of any typical QCD
energy scale. Similarly, the typical relative size of the electromagnetic (EM)
breaking of isospin symmetry is given by the fine structure constant
$\alpha\simeq 0.007$. For those reasons we can reasonably state that, for
observables with a non-vanishing isospin symmetric part, isospin symmetry is a
good approximation of reality with an $\bigo(1\%)$ relative error.
\begin{table}[b]
    \begin{center}
	\begin{tabular}{|l|c|c|}
		\cline{2-3}
		\multicolumn{1}{c|}{~} & $u$ & $d$                  \\
		\hline
		Mass ($\unit{\!}{\mega\eV}$) \citep{PDG2012}& 
        $2.3\left(^{+0.7}_{-0.5}\right)$ & 
		$4.8\left(^{+0.7}_{-0.3}\right)$                    \\
		Charge & $\frac{2}{3}e$ & $-\frac{1}{3}e$           \\	
		\hline
	\end{tabular}
    \end{center}
    \caption{Physical properties of the up and down quarks.}
    \label{tab:udquarks}
\end{table}

Nevertheless, it is interesting to notice that these little isospin breaking
corrections are crucial to describe the structure of atomic matter in the
Universe. Indeed, one particular effect of isospin symmetry breaking is the
mass splitting between the proton ($p$) and the neutron ($n$). This mass
difference is known experimentally with an impressive accuracy \citep{PDG2012}:
\begin{equation}
    \DM_N = M_p-M_n = \unit{-1.2933322(4)}{\mega\eV}
\end{equation}
The sign of this splitting makes the proton, and thus the hydrogen atom, a
stable physical state. Also, the size of $\DM_N$ determine the phase space
volume for the neutron $\beta$-decay $n\to p+e^-+\bar{\nu}_e$. At early times
of the Universe ($t\sim\unit{1}{\second}$ and $T\sim\unit{1}{\mega\eV}$) and
under standard assumptions\footnote{The neutrino number density
$n_{\nu}/n_{\gamma}$ is assumed to have the order of the baryon density number
which is very small. This assumption is not valid anymore in some new physics
scenarios but even in these hypothetical cases $n_n/n_p$ depends strongly on
$\DM_N$.}, the existence of $\beta$-decay allows to infer that the ratio of the
number of neutrons and protons is approximatively equal to:
\begin{equation}
    \frac{n_n}{n_p}\simeq\exp\left(\frac{\DM_N}{T}\right)
\end{equation}
This ratio is one important initial condition of Big Bang Nucleosynthesis.
Also, in our actual Universe, $\beta$-decay and its inverse process are known
to be responsible for the generation of a large majority of the stable nuclides
chart though nuclear transmutation. Even if the nucleon isospin mass splitting
is a well known quantity, predicting it from first principles is still an open
problem because of the complex non-perturbative interactions of quarks inside
the nucleon. The proton carries an additional EM self-energy compared to the
neutron, so just from QED one would expect to have $\DM_N>0$. However, the fact
that the experimental value of $\DM_N$ has the opposite sign suggests that the
strong isospin breaking effects are competing against the EM effects with a
larger magnitude. This would mean that an important part of the structure of
nuclear matter as we know it relies on a subtle cancellation between the small
EM and strong breaking effects of isospin symmetry in the nucleon system.
Therefore, it is fundamental to have a theoretical understanding of the nucleon
isospin mass splitting.

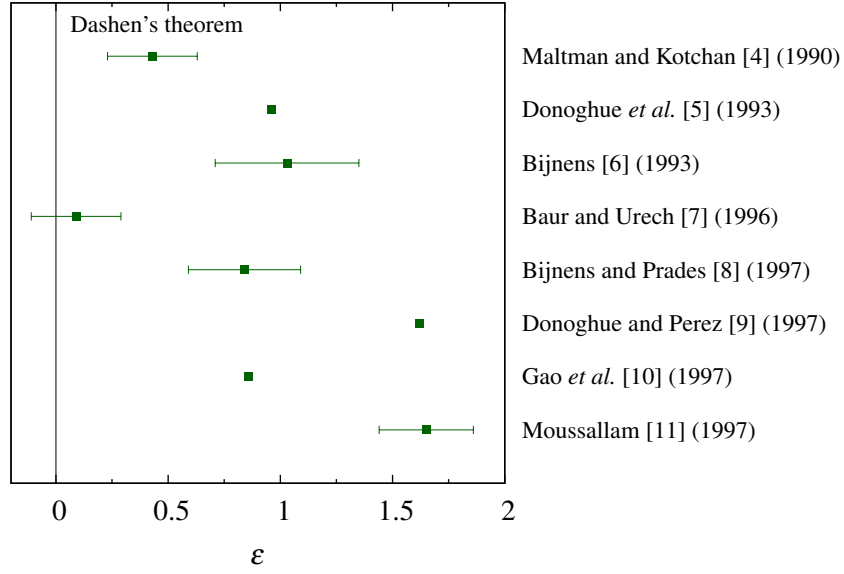
\begin{figure}[h!]
    \centering
        \input{figures/gp_comp_eps_phen_over}
    \caption{Summary of the analytical calculations of Dashen's theorem
    violation, $\epsilon$, defined in \eqref{eq:eps}. Results are presented in
    chronological order.}
    \label{fig:epsphen}
\end{figure}

Considering that isospin breaking effects in the hadron mass spectrum are
generally measured quite precisely, it is also interesting to understand how
one can use this information to deduce the masses of the individual $u$ and $d$
quark masses. For example, it is important to know if $m_u=0$ could be a
realistic solution to the strong CP problem. While recently (\cf the FLAG review
\citep{FLAG2011}) considerable progress has been made in determining
precisely the average up-down quark mass $m_{ud}$ from first principles, such
a computation is still missing for the individual masses. Because the kaon
($K$) is a pseudo-Goldstone boson of chiral symmetry breaking, the isospin mass
splitting $\DM^2_K=M_{K^+}^2-M_{K^0}^2$ is very sensitive to $\dm$. But in
order to extract $\dm$, one has to understand how to subtract the EM
contribution to this splitting. One well known result in this direction is
Dashen's theorem \citep{Dashen:1969tn} which states that, in the $\SU(3)$ chiral
limit, the EM Kaon splitting is equal to the EM pion ($\pi$) splitting:
\begin{equation}
    \DQEDM^2_K=\DQEDM^2_{\pi}+\bigo(\alpha m_s)
    \label{eq:dashth}
\end{equation}
This result is important because it is known that with good accuracy
(\cf\citep{FLAG2011}), $\DQEDM^2_{\pi}\simeq\DM^2_{\pi}$. The remaining
question is: how large are the $\bigo(\alpha m_s)$ corrections in
\eqref{eq:dashth}? One way to quantify these corrections is to consider the
dimensionless quantity $\epsilon$ defined in \citep{FLAG2011} as follows:
\begin{equation}
    \epsilon=\frac{\DQEDM^2_K-\DQEDM^2_{\pi}}{\DM^2_{\pi}}
    \label{eq:eps}
\end{equation}
This number is constructed such that it vanishes in the $\SU(3)$ chiral limit.
There were several attempts in the 1990s to compute these corrections
analytically from effective theories, they are illustrated in
\figref{fig:epsphen}. From these results, two conclusions can be made. First,
there is a clear disagreement between the different determinations from
effective theories which is an important motivation toward an \emph{ab-initio}
calculation of these corrections. Second, apart from this controversy it is
interesting to notice that most of these results indicate rather large
violations of Dashen's theorem. If this is the case, it means that
corrections to Dashen's theorem have to be known and taken into account in
order to compute the individual light quark masses.

The problems presented in this section, and more generally in any computation
of isospin corrections to low-energy QCD observables, are difficult to solve
because of the highly non-pertubative behavior of the strong interaction in
this regime. It has been shown recently (\cf for example
\citep{Durr:2008ir,Durr:2011iq}) that it is now possible to predict fundamental
isospin symmetric QCD observables through lattice QCD simulations with a full
control over the method's uncertainties. It is then reasonable to think that
lattice simulations could be a reliable way to understand and compute isospin
breaking effects. Moreover, besides the physical interest of these effects,
actual lattice calculations are reaching a sub-percent precision on several
standard observables and the assumption of isospin symmetry is becoming the
dominant source of systematic uncertainty.
\section{Lattice QCD and QED}
In this section we summarise lattice QCD simulation techniques and explain how
to include EM interactions in order to compute isospin breaking effects.
\subsection{Lattice QCD}
Let us consider a QCD observable $O$ in Euclidean space-time. Once one
integrates the quark fields, the expectation value of $O$ is given by the
following functional integral:
\begin{equation}
    \braket{O}=\frac{1}{\mathscr{Z}}\int\fdiff G_{\mu}\,
    O^{\mathrm{Wick}}[G_{\mu},(\slashed{D}+M)^{-1}]\det(\slashed{D}+M)
    e^{-S_{\mathrm{YM}}[G_{\mu}]}
\end{equation}
where $\mathscr{Z}$ is the partition function, $G_{\mu}$ is the gluon field,
$O^{\mathrm{Wick}}$ is the gluonic observable obtained from $O$ once the quark
fields have been Wick contracted, $D_{\mu}$ is the gauge covariant derivative,
$M$ is the quark mass matrix and $S_{\mathrm{YM}}$ is the pure gauge Yang-Mills
action. A remarkable feature of this integral is that
$\frac{1}{\mathscr{Z}}\det(\slashed{D}+M)e^{-S_{\mathrm{YM}}[G_{\mu}]} \fdiff
G_{\mu}$ is a probability density. Therefore, if one is able to draw an large
number of gluon field according to this law, the expectation value $\braket{O}$
can be evaluated numerically through a Monte-Carlo estimation. This idea is the
essence of lattice QCD calculations.

To perform such Monte-Carlo calculations, it is of course needed to discretise
QCD. To do this, we reduce space-time to a 4-dimensional periodic hypercubic
lattice with a lattice spacing $a$. In order to discretise the gauge action
while keeping exact gauge invariance, it is known that one needs to consider
the gauge links $U_{\mu}=e^{iG_{\mu}}$ as fundamental variables of the theory.
Writing an action for those link variables that converge to the Yang-Mills
action in the continuum limit $a\to 0$ does not present any particular
difficulties that require comment here. However, discretising the Dirac action
is a non-trivial problem. Indeed, a naive discretisation of Dirac action leads
to a theory where any fermionic flavour exist with a multiplicity of $16$. This
is the so-called \emph{doubler problem}. This issue is deeply related to the
difficulty of reconstructing chiral symmetry on a discrete space-time, a
detailed description of this problem and its possible solutions can be found in
\citep{Kaplan:2011wx}.
\subsection{QED in finite volume}
\label{ssec:fvol}
In order to include QED in lattice simulations, one needs to understand how to
formulate electrodynamics consistently in a finite volume. Compared to QCD, the
main difficulty comes from the absence of a mass gap in the theory:
physical QED states can have an energy arbitrary close to $0$. This will have
two important consequences. First, the propagation of gauge modes is not
suppressed by any mass scale: the interaction is long-ranged and important
finite volume (FV) effects are expected. Second, momentum quantisation in
finite volume will alter the structure of infrared (IR) divergences in the
theory.

We begin by discussing IR divergences. Let us consider a diagrammatic
contribution $\mathscr{D}$ to a correlation function featuring a photon loop
(\eg the 1-loop part of the electron EM self-energy). In infinite volume,
$\mathscr{D}$ will have the following form:
\begin{equation}
    \mathscr{D}=\int\frac{\diff^4 k}{(2\pi)^4}\frac{1}{k^2}f(k,p_1,\dots,p_n)
    \label{eq:irdiv}
\end{equation}
where the $p_j$ are the external momenta of the diagram. We assume that we are
in the non-trivial situation where the degree of $f(k,p_1,\dots,p_n)$ in $k$
for $k^2\to0$ is strictly less than $2$. In this context, the integrand in
\eqref{eq:irdiv} diverges for $k^2\to 0$. Nevertheless, depending on the
behaviour of $f$ in the vicinity of $k=0$ and the value of the external
momenta, this divergence can be integrable and $\mathscr{D}$ can be finite. It
is well known in infinite volume QED that IR divergences do not affect
physical observables. In finite volume, momenta are quantised and the integral
in \eqref{eq:irdiv} becomes a sum over discrete momentum space which depends of
the choice of boundary condition for the photon field. For periodic boundary
conditions, $k=0$ is part of momentum space\footnote{With periodic boundary
conditions, momentum is quantised in the form
$k_{\mu}=\frac{2\pi}{L_{\mu}}n_{\mu}$ where $L_{\mu}$ is the size of the volume
in direction $\mu$ and the $n_{\mu}$ are integers.} and the FV version of
$\mathscr{D}$ is infinite independently of the properties of $f$ in the
vicinity of $k=0$. This shows that a naive formulation of FV QED can easily
lead to a divergent theory. Therefore, it is needed to provide an appropriate
IR regularisation of the FV theory. Here we present a possible solution which
allows to conserve gauge invariance (\ie a massless photon). We chose to use
periodic boundary conditions and remove the zero mode of the gauge field from
the path integral variables. This will subtract the infinite term in the
momentum sum in the FV version of $\mathscr{D}$. Also, the gauge field is
modified on a set of measure $0$ so this alteration disappear in the infinite
volume limit, where the momentum sum becomes an integral. This choice of
regularisation is the one adopted by all the actual lattice QCD and QED
projects. A more physical description of this regularisation can be found in
\citep{Portelli:2010tz}.

The other difficulty comes from the infinite range of the EM interaction:
important FV effects are expected. These effects are generated by photons going
around the finite, periodic space-time. Formally, these effects are given by
the difference between the momentum integral and sum in contributions of the
form \eqref{eq:irdiv}. These effects have been worked out in partially quenched
chiral perturbation theory coupled to photons (PQ$\chi$PT+QED) in
\citep{Hayakawa:2008ci}, in the approximation where the time dimension is
infinite. In \citep{deDivitiis:2013vc}, an asymptotic expansion of these
results is performed, and it is shown that the leading order FV effects are
$\bigo(L^{-1})$ where $L$ is the length of spatial dimensions. However, it has
been observed in \citep{Blum:2010ge,Portelli:2012vz} that this PQ$\chi$PT+QED
model does not reproduce well the FV effects observed in the data.
Nevertheless, a $\bigo(L^{-1})$ volume dependence is observed in the results of
\citep{Borsanyi:2013va} using a wide range of volumes. This confirms the
intuition of large FV QED effects. Indeed, the FV effects from QCD are known
(\cf for example \citep{Colangelo:2005gx}) to decay exponentially with
$M_{\pi}L$ and are quickly dominated by the power-like behaviour of the QED FV
effects. To conclude, it seems crucial to use large volumes (above
$\sim(\unit{6}{\fm})^3$) in lattice simulations in order to control completely
the infinite volume limit.

\subsection{Inclusion of QED in lattice QCD simulations}
\label{ssec:qcdqed}
Up to now, there are essentially two ways of including the EM interactions in
lattice calculations. First, similarly to the traditional lattice QCD methods,
one can include QED stochastically by generating random EM fields in order to
obtain a Monte-Carlo estimation of the path integral. Also, as it was done by
the RM123 collaboration in \citep{deDivitiis:2013vc}, the correlation functions
of QCD and QED can be expanded perturbatively in $\alpha$. By doing this, the
QED corrections can be calculated from expectation values of pure QCD operators.

First of all, one needs to formulate QED on a finite, periodic lattice. Up to
now, the formulation commonly used is the so-called \emph{non-compact lattice
QED}. The gauge action of this theory is just a naive discretisation of Maxwell
action:
\begin{equation}
    S_{\mathrm{Maxwell}}[A_{\mu}]
    =\frac{a^4}{4}\sum_{\mu,\nu,x}[\partial_{\mu}A_{\nu}(x)-
    \partial_{\nu}A_{\mu}(x)]^2
\end{equation}
where $A_{\mu}$ is the real EM potential and $\partial_{\mu}$ is the forward 
finite difference in direction $\mu$:
\begin{equation}
    \partial_{\mu}f(x)=\frac{1}{a}[f(x+a\hat{\mu})-f(x)]
\end{equation}
where $\hat{\mu}$ is the unitary vector in direction $\mu$. Thanks to the
abelianity of the $\U(1)$ group, this action remains gauge invariant even on a
discrete space-time. Also, accordingly to the discussion in \secref{ssec:fvol},
one must use a FV IR regularisation such as the subtraction of $A_{\mu}$ zero
mode. Then this EM gauge field can be coupled to the lattice quarks in the
following way: one first constructs the $\U(1)$ gauge links
$U^{\QED}_{\mu}=e^{ieQA_{\mu}}$ where $Q$ is the charge matrix in flavour
space. Then the links which are used in the discrete version of the gauge
covariant Dirac operator are the $U_{\mu}=U^{\QED}_{\mu}U^{\QCD}_{\mu}$, where
the $U^{\QCD}_{\mu}$ are the lattice QCD link variables.

Let us now discuss how to stochastically simulate QCD and QED. One considerable
advantage of pure gauge QED over QCD is that it is a free theory: there is no
photon-photon interaction. In other words, the Maxwell action is quadratic and
$e^{-S_{\mathrm{Maxwell}}[A_{\mu}]}$ is just a Gaussian weight. So drawing pure
gauge EM fields is a trivial matter of drawing Gaussian random numbers. There
is no need to go through a Markov chain simulation as it is usually done in
QCD. This simple Gaussian random drawing can then be easily included in any
Markovian process used to simulate QCD (generally a variant of the Hybrid
Monte-Carlo algorithm). However, the distribution of the QCD and QED fields in
the path integral are not independent: they are mixed inside the fermionic
determinant. It means that in principle one must generate simultaneously QCD
and QED fields. As the generation of QCD configurations represents generally
the large majority of the computational cost in lattice simulations, almost
every calculations up to now used the \emph{electro-quenched} approximation
where one neglects the QED contribution to the determinant. This allows to
reuse previously generated QCD configurations. Physically, this approximation
is equivalent to assume that sea quarks are neutral. We will discuss later the
precision of this approximation. It is possible to make a full QCD and QED
stochastic calculation reusing previously generated QCD configurations by
computing \emph{a posteriori} the needed correction to the fermionic
determinant: this is the so-called \emph{reweighting} method. This method has
been used in \citep{Duncan:2005cq,Ishikawa:2012iw,Aoki:2012uu}. Although
interesting, the use of reweighting generally degrades the signal its
computational cost increases very quickly with the lattice volume. This happens
because of the non-local nature of the determinant. Therefore, if one considers
the need to use large volumes to control EM FV effects, reweighting looks like
a limited option to include QED sea effects.

As explained in the introduction of this section, a second option to include
QED in lattice calculations is to do it in a perturbation theory fashion. If
one considers, for example, the lattice 2-point function of some local spin $0$
operator $O$ in QCD and QED, it is easy to show that:
\begin{equation}
    \braket{O(x)O(0)^{\dagger}}_{\QCD+\QED}=
    \braket{O(x)O(0)^{\dagger}}_{\QCD}+e^2\sum_{y,z,\mu,\nu}
    D_{\mu\nu}(y-z)\braket{O(x)
    \timeor[J_{\mu}(y)J_{\nu}(z)]O(0)^{\dagger}}_{\QCD}+\bigo(e^4)
    \label{eq:pertexp}
\end{equation}
where $J_{\mu}$ is the conserved EM current, $D_{\mu\nu}$ is the lattice free
photon propagator and the indices on expectation values indicate in which
theory the path integral is considered. Of course the definition of $J_{\mu}$
depends on the choice of lattice action for the quarks. With this expansion,
the first order EM corrections to $O$ propagator can be computed as a pure QCD
4-point function where one considers all possible insertions of two EM
currents. This method is used and presented in detail in
\citep{deDivitiis:2013vc}. The authors of this paper also show that the
$\bigo(\delta m)$ isospin corrections can be computed with the same kind of
perturbative expansion. Also, in this work were neglected the quark
disconnected diagrams featuring EM current self-contractions. This type of
diagram is known to be difficult and expensive to compute in lattice QCD. This
approximation is exactly equivalent to the electro-quenched approximation
mentioned before. Technically, the computation of the $4$-point function in
\eqref{eq:pertexp} is much more complicated than the direct stochastic
estimation, described before, of the QCD and QED 2-point function.
Nevertheless, this perturbative expansion is a relevant method to obtain a
specific diagrammatic contribution to a correlation function (\eg the
computation of hadronic light-by-light contribution to the muon anomalous
magnetic moment).

Let us finish this section by discussing the precision of the electro-quenched
approximation. In this context, the leading missing contributions are quark
loops with one photon emission and an arbitrary number of gluon emissions.
These contributions will always have the form $\tr(QS)$ in flavour space where
$Q$ is the quark charge matrix and $S$ is the full QCD quark propagator. In the
three flavour theory (considering $u$,$d$ and $s$ flavours), $Q$ is traceless.
So if $S$ is proportional to the unit matrix (\ie if $m_u=m_d=m_s$), $\tr(QS)$
vanishes. This means that the sea EM contributions are $\SU(3)$ flavour
symmetry breaking effects. Also, it is easy to show that these contributions
are suppressed by the number of colours $N_c=3$. Building $\SU(3)$ flavour
breaking ratios from the hadron spectrum (\ie $(M_{\Sigma}-M_N)/M_N$) one can
observe that $\SU(3)$ breaking effect are typically $\bigo(30\%)$ corrections.
Dividing this number by $N_c$, one can then estimate that the electro-quenched
approximation is correct up to $\bigo(10\%)$ corrections.

\subsection{Summary of actual lattice QCD and QED calculations}
In \tabref{tab:collab} we summarise the parameters of lattice QCD and QED
projects. In addition to this table, we should mention that most of these works
are based on techniques that were developed for the first time in the
pioneering work from A. Duncan \textit{et al.}
\citep{Duncan:1996cy,Duncan:1997cp}. Also there are some works from NPLQCD
\citep{Beane:2007eh}, UKQCD-QCDSF \citep{Horsley:2012ue} and P.E. Shanahan
\textit{et al.} \citep{Shanahan:2013kg} which consider only strong isospin
corrections in lattice QCD. Regarding analytical calculations with effective
theories, beside the references cited in \figref{fig:epsphen}, there is a
calculation of the EM mass splitting in the baryon octet in the review
\citep{Gasser:1982ic} and a determination of the EM mass splitting of the
nucleon in \citep{WalkerLoud:2012gq}.
\begin{table}
    \begin{center}
    \begin{tabular}{|r|c|c|c|c|c|}
        \hline
        collaboration       & RBC-UKQCD & MILC   & BMWc & PACS-CS & RM123\\
        references & \citep{Blum:2007gh,Blum:2010ge} & 
        \citep{Basak:2008td,Basak:2010ww,Basak:2012zx,Basak:2013ug} & 
        \citep{Portelli:2010tz,Portelli:2012vz,Borsanyi:2013va,BMW:2013} &
        \citep{Aoki:2012uu}&
        \citep{deDivitiis:2011wf,deDivitiis:2013vc} \\
        \hline
        fermion action      & DW     & Asqtad & 2-HEX tlSW & npSW  & tmWil \\
        $N_f$               & $2+1$     & $2+1$  & $2+1$  & $1+1+1$ & $2$  \\
        QED type            & qQED    & qQED   & qQED   & QED     & qpQED\\
        $N_{\mathrm{sim.}}$ & $7$       & $7$    & $39$   & $1$     & $13$ \\
        $\min(M_{\pi})~(\mega\eV)$  & $250$     & $233$  & $120$  & $135$  
            & $270$\\
        $a~(\fm)$           & $0.11$    & $0.06\,\text{--}\, 0.12$ & 
            $0.05\,\text{--}\, 0.12$ & $0.09$ & $0.05\,\text{--}\, 0.10$\\
        $N_{a}$             & $1$       & $3$    & $5$    & $1$     & $4$  \\
        $L~(\fm)$    & $1.8\,\text{--}\, 2.7$ & $2.4\,\text{--}\, 3.6$ &  
            $1.9\,\text{--}\, 6$ & $2.9$ & $1.6\,\text{--}\, 2.6$\\
        $N_{\mathrm{vol.}}$ & $2$       & $5$    & $17$   & $1$     & $6$  \\
        \hline
    \end{tabular}
    \end{center}
    \caption{Summary of the main lattice QCD and QED calculations. The
    first line is the fermion action used, the abbreviations are
    from \citep[p. 52]{FLAG2011}. The second line is the number of flavours used
    in the gauge configuration generation. The third line gives the method used
    to implement QED: ``q'' means that the electro-quenched approximation is
    used and ``p'' means that the perturbative method described in the previous
    section is used. The fourth line gives the number of simulation points used.
    The fifth line indicates the minimum pion mass reached. The sixth line is
    the range of lattice spacing used and the seventh line indicates their
    number. Similarly, the eighth and ninth lines are respectively the range and
    the number of lattice spatial extents used.}
    \label{tab:collab}
\end{table}
\section{Dashen's theorem and light quark masses}
In order to compute corrections to Dashen's theorem, one has to compute the QED
contribution to the pion and kaon mass splittings. Let us write down the
leading order isospin corrections to the squared mass splitting of a
pseudoscalar $P$:
\begin{equation}
    \DM^2_P=M_{P^+}^2-M_{P^0}^2=\alpha A_P + \dm B_P
\end{equation}
where $A_P$ and $B_P$ are two functions of the isospin symmetric parameters of
the theory. Then it is tempting to define:
\begin{equation}
    \DQEDM^2_P=\alpha A_P\quad\text{and}\quad\DQCDM^2_P=\dm B_P
\end{equation}
These definitions are ambiguous because $\alpha$, $m_u$ and $m_d$ depend on
each other through renormalisation. More explicitly, $\DQEDM^2_P$ also depends
on $\dm$ and $\DQCDM^2_P$ does on $\alpha$. So in principle, one has to specify
a scheme (and a scale) to separate QED and QCD isospin breaking effects. Let us
estimate the typical size of the associated scheme ambiguity. If we assume that
one determined the QCD splitting of $P$ for $\alpha=0$, how does this result
differ from a determination at the physical value of $\alpha$? Each individual
$u$ and $d$ quark mass will differ by a relative $\bigo(\alpha)$ contribution,
yielding:
\begin{equation}
    \DQCDM^2_P(\alpha)=\DQCDM^2_P(0)+\bigo(\alpha\dm,\alpha m_{ud})
\end{equation}
As we are only interested in leading order isospin corrections, we can ignore
the $\bigo(\alpha\dm)$ contribution. At the physical quark masses, it is usual
and reasonable to consider that $\bigo(m_{ud})=\bigo(\dm)$. Including this
additional assumption in our power counting, we can consider that $\bigo(\alpha
m_{ud})$ corrections are negligible. A similar argument shows that the QED mass
splittings ambiguity is also of order $\bigo(\alpha\dm,\alpha m_{ud})$. To
conclude, we showed that the separation of QED and QCD contributions to isospin
breaking effects can be obtained unambiguously, at leading order, at the
physical quark masses and up to small $\bigo(\alpha m_{ud})$ corrections.

\begin{figure}[ph!]
    \centering
        \input{figures/gp_comp_eps_lat_over}
        \vspace{0.1cm}
    \caption{Summary of lattice calculations of Dashen's theorem violation,
    $\epsilon$, defined in \eqref{eq:eps}. Results are presented in
    chronological order. A blue error bar represents the statistical error and
    a red one the systematic error. The RBC-UKQCD paper \citep{Blum:2007gh}
    gives two different results corresponding to two cuts in their dataset. We
    decided to interpret this interval as a systematic error. Most of these
    papers do not quote explicitly $\epsilon$ as defined in \eqref{eq:eps} and
    the FLAG review \citep{FLAG2011}. When missing, we computed the value of
    $\epsilon$ only using information from the relevant paper, the PDG review
    \citep{PDG2012} and the FLAG review \citep{FLAG2011}. The results in italic
    come from conference proceedings}
    \label{fig:epslat}
\end{figure}
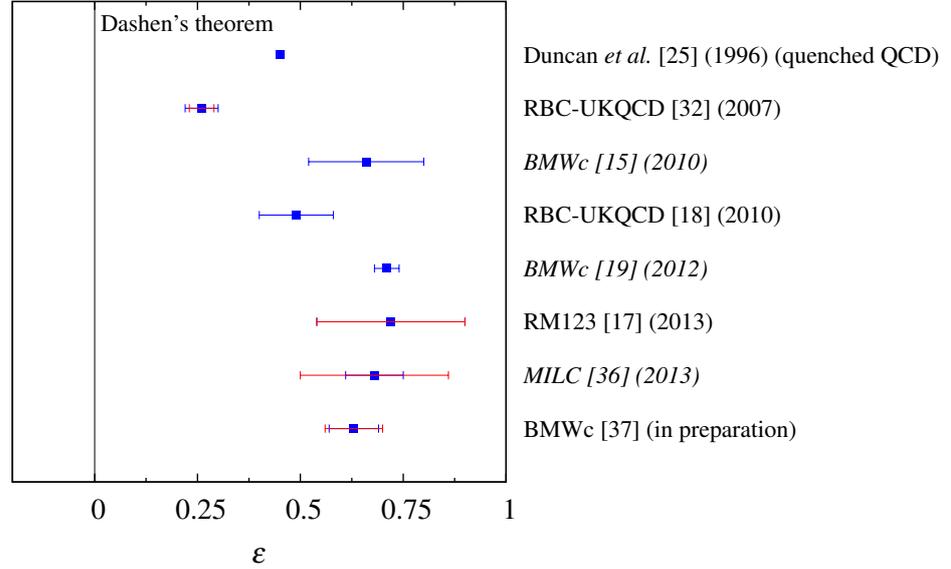

Now that we have clarified how to separate QED and QCD isospin breaking
effects, let us review the actual lattice results concerning the corrections to
Dashen's theorem. These results are summarised in figure \figref{fig:epslat}. If
one compares these results to the phenomenological ones presented in
\figref{fig:epsphen}, it is clear that the lattice results are really helping to
give a consistent answer to the problem of Dashen's theorem corrections. In
2011, the last FLAG review \citep{FLAG2011} quoted $\epsilon=0.7(5)$ as a
global result for Dashen's theorem corrections. Considering the results
obtained since 2012, especially \citep{deDivitiis:2013vc,BMW:2013}, lattice
calculations reduce significantly this uncertainty. So we can say that lattice
QCD and QED are on the way to solve the long-lasting controversy around
Dashen's theorem. Nevertheless, $\epsilon$ is computed from QED splittings and
all the results up to now have been calculated in the electro-quenched
approximation, which, as discussed in \secref{ssec:qcdqed}, implies an
$\bigo(10\%)$ uncontrolled uncertainty on QED mass splittings.

\begin{figure}[ph!]
    \centering
        \input{figures/gp_comp_mq_lat_over}
        \vspace{0.1cm}
    \caption{Summary of lattice calculations of the up and down quark mass
    ratio. Graphical conventions are the same than in \figref{fig:epslat}.}
    \label{fig:mqlat}
\end{figure}
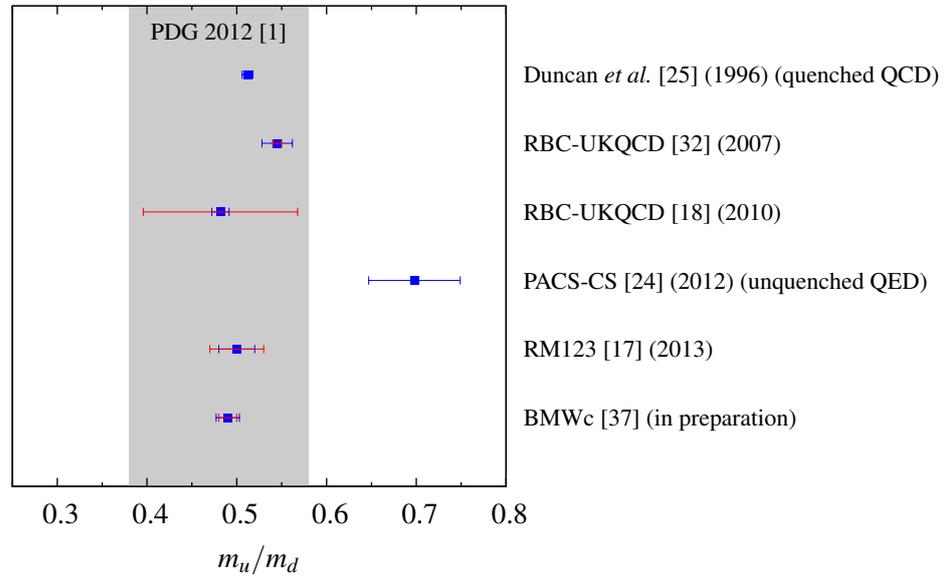

In \figref{fig:mqlat} we summarise the actual lattice calculation of the light
quark mass ratio $m_u/m_d$. It is important to notice that, even with QED taken
into account, this ratio is still independent of the renormalisation scheme up
to higher order isospin corrections. If one compares the most recent lattice
results and the PDG review average illustrated by the grey band in
\figref{fig:mqlat}, it is clear that the inclusion of isospin breaking effect in
lattice calculations greatly improves the determination of this ratio. We are
now in a situation where we can state that $m_u\neq 0$ with a significance of
the order of $10$ standard deviations. This strongly excludes $m_u=0$ as a
possible explanation for the absence of CP violation in the strong interaction.
\section{Octet baryon isospin mass splittings}
Compared to meson masses, baryon masses are more difficult to compute precisely
through lattice computation. Because of this, there are few works on the
computation of the isospin breaking corrections to baryon masses. Most of the
actual work is focused on the nucleon splitting. These results are summarised
the results in \figref{fig:Nall}. Since a number of projects only determine the
QCD or QED contribution to this splitting, we choose to summarise results for
each of these contributions rather than for the full splitting. Even if most of
these results have still a small significance, they suggest that the nucleon
splitting, whose value is crucial for nuclear matter stability, comes from a
subtle cancelation between QED and QCD isospin breaking effects. Concerning the
baryon octet, there are two pure QCD studies
\citep{Horsley:2012ue,Shanahan:2013kg}. Because QED was not taken into account
in these two projects, they both had to use imprecise phenomenological inputs
related to the light quark mass ratio. Beside the early work
\citep{Duncan:1997cp}, which was performed in quenched QCD, the recent results
from BMWc \citep{Borsanyi:2013va} constitute the only complete determination of
QCD and QED isospin corrections to the baryon octet. Their results are
summarised in \figref{fig:bmwsplit}.
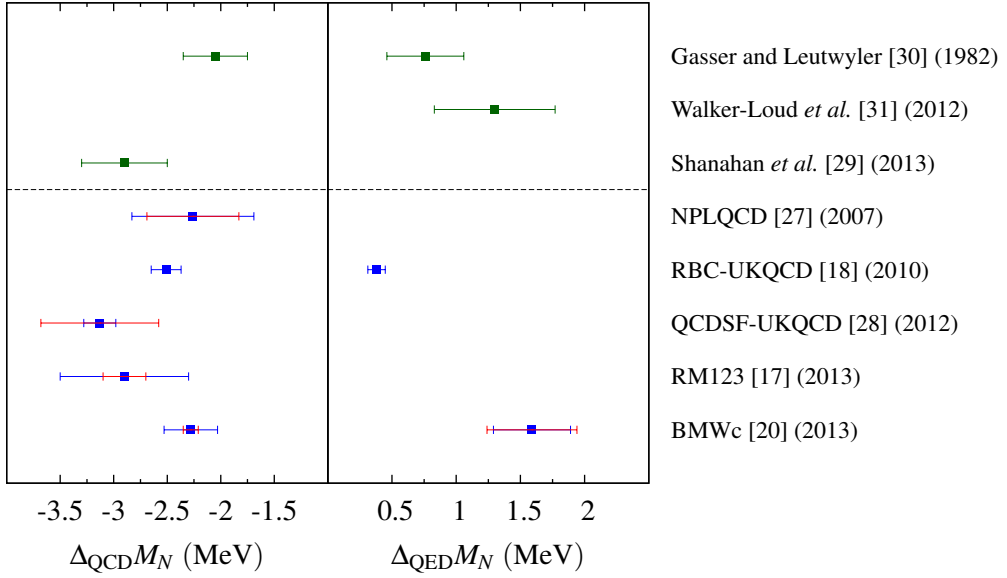
\begin{figure}[h]
    \centering
        \input{figures/gp_comp_N_all_over}
        \vspace{0.2cm}
    \caption{Summary of calculations of the QCD and QED contributions to the
    nucleon mass splitting $\DM_N=M_p-M_n$. As in previous plots, green points
    are phenomenological results and blue/red points are lattice results.}
    \label{fig:Nall}
\end{figure}
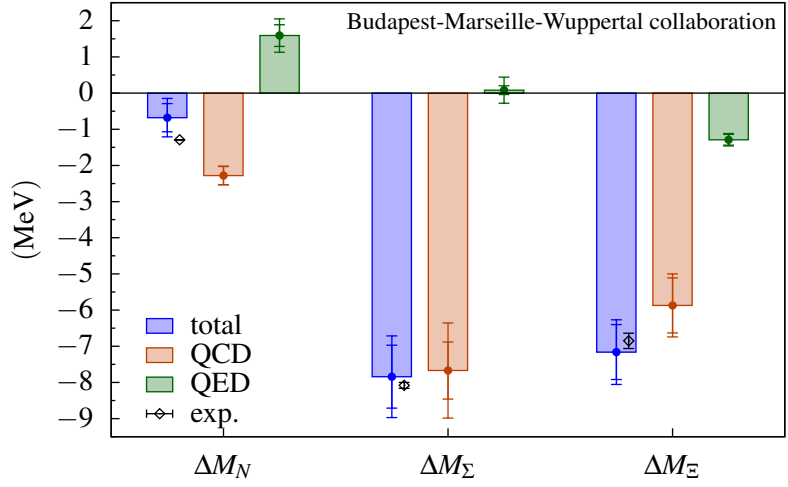
\begin{figure}[h]
    \centering
        \input{figures/gp_final_split_over}
    \caption{Summary of BMWc results \citep{Borsanyi:2013va} for the isospin
    mass splittings of the octet baryons. Also shown are the individual QCD and
    QED contributions to these splittings. The bands indicate the size of the
    splittings and contributions. On the points, the error bars are the
    statistical and total uncertainties (statistical and systematic combined in
    quadrature). For comparison, the experimental values for the total
    splittings are also displayed as black points.}
    \label{fig:bmwsplit}
\end{figure}
\section{Conclusion and perspectives}
The first lattice studies of isospin breaking effects
presented in this review show that lattice simulations are encouragingly
entering the era where the computation of such small effects is possible.

It is clear that it is now time to go beyond the electro-quenched
approximation. This will allow to produce fully controlled high precision
predictions for corrections to Dashen's theorem and for individual up and down
quark masses. This is also needed in order to compute the nucleon splitting
with a high level of significance, which would constitute an \emph{ab-initio}
proof of nuclear matter stability.

Also one can think about computing more sophisticated isospin corrections such
as, for example, corrections to decay constants and hadronic matrix elements.
This would require theoretical work to define properly these quantities in QCD
and QED (\cf the interesting discussion in \citep{Gasser:2010ee}). Once this is
achieved, lattice calculations will be able to fully simulate the Standard
Model in the low energy regime.

\acknowledgments{I would like to gratefully thank Laurent Lellouch for his
help and support along the production of this review and Vittorio Lubicz and
Chris Sachrajda for helpful discussions.}
\bibliographystyle{pos}
\bibliography{article}

\end{document}

%% file: figures/gp_comp_eps_phen_over.tex
\begingroup
  \makeatletter
  \providecommand\color[2][]{%
    \GenericError{(gnuplot) \space\space\space\@spaces}{%
      Package color not loaded in conjunction with
      terminal option `colourtext'%
    }{See the gnuplot documentation for explanation.%
    }{Either use 'blacktext' in gnuplot or load the package
      color.sty in LaTeX.}%
    \renewcommand\color[2][]{}%
  }%
  \providecommand\includegraphics[2][]{%
    \GenericError{(gnuplot) \space\space\space\@spaces}{%
      Package graphicx or graphics not loaded%
    }{See the gnuplot documentation for explanation.%
    }{The gnuplot epslatex terminal needs graphicx.sty or graphics.sty.}%
    \renewcommand\includegraphics[2][]{}%
  }%
  \providecommand\rotatebox[2]{#2}%
  \@ifundefined{ifGPcolor}{%
    \newif\ifGPcolor
    \GPcolortrue
  }{}%
  \@ifundefined{ifGPblacktext}{%
    \newif\ifGPblacktext
    \GPblacktexttrue
  }{}%
  \let\gplgaddtomacro\g@addto@macro
  \gdef\gplbacktext{}%
  \gdef\gplfronttext{}%
  \makeatother
  \ifGPblacktext
    \def\colorrgb#1{}%
    \def\colorgray#1{}%
  \else
    \ifGPcolor
      \def\colorrgb#1{\color[rgb]{#1}}%
      \def\colorgray#1{\color[gray]{#1}}%
      \expandafter\def\csname LTw\endcsname{\color{white}}%
      \expandafter\def\csname LTb\endcsname{\color{black}}%
      \expandafter\def\csname LTa\endcsname{\color{black}}%
      \expandafter\def\csname LT0\endcsname{\color[rgb]{1,0,0}}%
      \expandafter\def\csname LT1\endcsname{\color[rgb]{0,1,0}}%
      \expandafter\def\csname LT2\endcsname{\color[rgb]{0,0,1}}%
      \expandafter\def\csname LT3\endcsname{\color[rgb]{1,0,1}}%
      \expandafter\def\csname LT4\endcsname{\color[rgb]{0,1,1}}%
      \expandafter\def\csname LT5\endcsname{\color[rgb]{1,1,0}}%
      \expandafter\def\csname LT6\endcsname{\color[rgb]{0,0,0}}%
      \expandafter\def\csname LT7\endcsname{\color[rgb]{1,0.3,0}}%
      \expandafter\def\csname LT8\endcsname{\color[rgb]{0.5,0.5,0.5}}%
    \else
      \def\colorrgb#1{\color{black}}%
      \def\colorgray#1{\color[gray]{#1}}%
      \expandafter\def\csname LTw\endcsname{\color{white}}%
      \expandafter\def\csname LTb\endcsname{\color{black}}%
      \expandafter\def\csname LTa\endcsname{\color{black}}%
      \expandafter\def\csname LT0\endcsname{\color{black}}%
      \expandafter\def\csname LT1\endcsname{\color{black}}%
      \expandafter\def\csname LT2\endcsname{\color{black}}%
      \expandafter\def\csname LT3\endcsname{\color{black}}%
      \expandafter\def\csname LT4\endcsname{\color{black}}%
      \expandafter\def\csname LT5\endcsname{\color{black}}%
      \expandafter\def\csname LT6\endcsname{\color{black}}%
      \expandafter\def\csname LT7\endcsname{\color{black}}%
      \expandafter\def\csname LT8\endcsname{\color{black}}%
    \fi
  \fi
  \setlength{\unitlength}{0.0500bp}%
  \begin{picture}(5668.00,4250.00)%
    \gplgaddtomacro\gplbacktext{%
      \csname LTb\endcsname%
      \put(618,205){\makebox(0,0){\strut{} 0}}%
      \put(1455,205){\makebox(0,0){\strut{} 0.5}}%
      \put(2292,205){\makebox(0,0){\strut{} 1}}%
      \put(3129,205){\makebox(0,0){\strut{} 1.5}}%
      \put(3966,205){\makebox(0,0){\strut{} 2}}%
      \put(2124,-125){\makebox(0,0){\strut{}$\varepsilon$}}%
    }%
    \gplgaddtomacro\gplfronttext{%
      \csname LTb\endcsname%
      \put(1371,3876){\makebox(0,0){\footnotesize Dashen's theorem}}%
    }%
    \gplgaddtomacro\gplbacktext{%
    }%
    \gplgaddtomacro\gplfronttext{%
      \csname LTb\endcsname%
      \put(4096,3635){\makebox(0,0)[l]{\strut{}\footnotesize{\citet{Maltman:1990dh}~(\citeyear{Maltman:1990dh})}}}%
      \put(4096,3234){\makebox(0,0)[l]{\strut{}\footnotesize{\citet{Donoghue:1993bm}~(\citeyear{Donoghue:1993bm})}}}%
      \put(4096,2832){\makebox(0,0)[l]{\strut{}\footnotesize{\citet{Bijnens:1993go}~(\citeyear{Bijnens:1993go})}}}%
      \put(4096,2431){\makebox(0,0)[l]{\strut{}\footnotesize{\citet{Baur:1996gf}~(\citeyear{Baur:1996gf})}}}%
      \put(4096,2030){\makebox(0,0)[l]{\strut{}\footnotesize{\citet{Bijnens:1997ku}~(\citeyear{Bijnens:1997ku})}}}%
      \put(4096,1629){\makebox(0,0)[l]{\strut{}\footnotesize{\citet{Donoghue:1997kt}~(\citeyear{Donoghue:1997kt})}}}%
      \put(4096,1227){\makebox(0,0)[l]{\strut{}\footnotesize{\citet{Gao:1997hq}~(\citeyear{Gao:1997hq})}}}%
      \put(4096,826){\makebox(0,0)[l]{\strut{}\footnotesize{\citet{Moussallam:1997jm}~(\citeyear{Moussallam:1997jm})}}}%
    }%
    \gplbacktext
    \put(0,0){\includegraphics{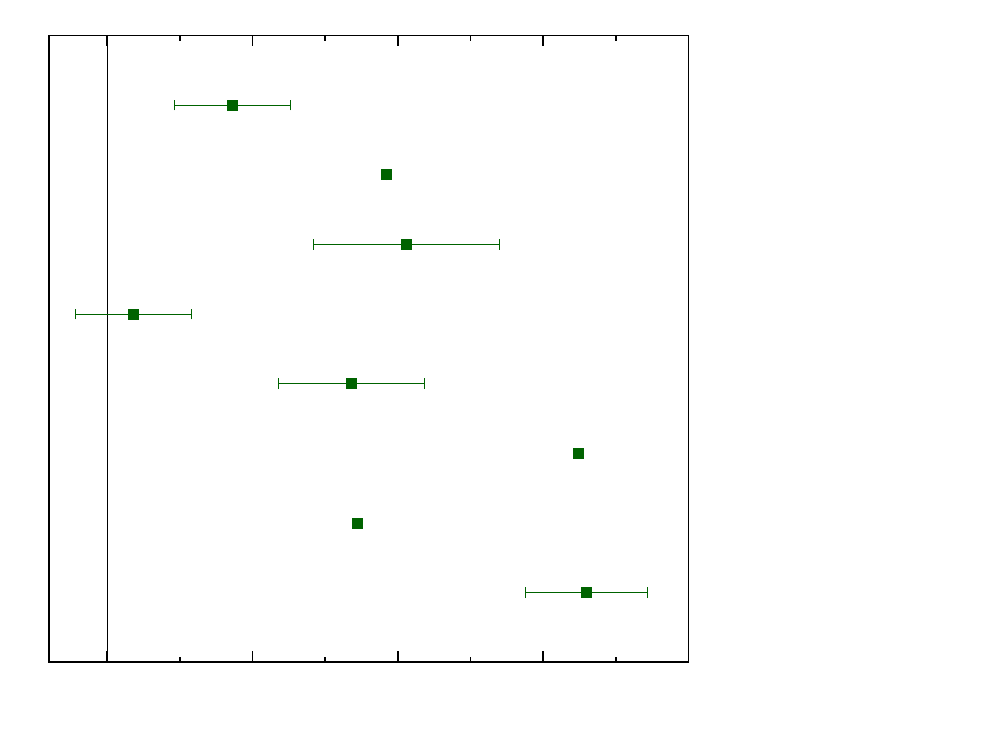}}%
    \gplfronttext
  \end{picture}%
\endgroup

%% file: figures/gp_comp_eps_lat_over.tex
\begingroup
  \makeatletter
  \providecommand\color[2][]{%
    \GenericError{(gnuplot) \space\space\space\@spaces}{%
      Package color not loaded in conjunction with
      terminal option `colourtext'%
    }{See the gnuplot documentation for explanation.%
    }{Either use 'blacktext' in gnuplot or load the package
      color.sty in LaTeX.}%
    \renewcommand\color[2][]{}%
  }%
  \providecommand\includegraphics[2][]{%
    \GenericError{(gnuplot) \space\space\space\@spaces}{%
      Package graphicx or graphics not loaded%
    }{See the gnuplot documentation for explanation.%
    }{The gnuplot epslatex terminal needs graphicx.sty or graphics.sty.}%
    \renewcommand\includegraphics[2][]{}%
  }%
  \providecommand\rotatebox[2]{#2}%
  \@ifundefined{ifGPcolor}{%
    \newif\ifGPcolor
    \GPcolortrue
  }{}%
  \@ifundefined{ifGPblacktext}{%
    \newif\ifGPblacktext
    \GPblacktexttrue
  }{}%
  \let\gplgaddtomacro\g@addto@macro
  \gdef\gplbacktext{}%
  \gdef\gplfronttext{}%
  \makeatother
  \ifGPblacktext
    \def\colorrgb#1{}%
    \def\colorgray#1{}%
  \else
    \ifGPcolor
      \def\colorrgb#1{\color[rgb]{#1}}%
      \def\colorgray#1{\color[gray]{#1}}%
      \expandafter\def\csname LTw\endcsname{\color{white}}%
      \expandafter\def\csname LTb\endcsname{\color{black}}%
      \expandafter\def\csname LTa\endcsname{\color{black}}%
      \expandafter\def\csname LT0\endcsname{\color[rgb]{1,0,0}}%
      \expandafter\def\csname LT1\endcsname{\color[rgb]{0,1,0}}%
      \expandafter\def\csname LT2\endcsname{\color[rgb]{0,0,1}}%
      \expandafter\def\csname LT3\endcsname{\color[rgb]{1,0,1}}%
      \expandafter\def\csname LT4\endcsname{\color[rgb]{0,1,1}}%
      \expandafter\def\csname LT5\endcsname{\color[rgb]{1,1,0}}%
      \expandafter\def\csname LT6\endcsname{\color[rgb]{0,0,0}}%
      \expandafter\def\csname LT7\endcsname{\color[rgb]{1,0.3,0}}%
      \expandafter\def\csname LT8\endcsname{\color[rgb]{0.5,0.5,0.5}}%
    \else
      \def\colorrgb#1{\color{black}}%
      \def\colorgray#1{\color[gray]{#1}}%
      \expandafter\def\csname LTw\endcsname{\color{white}}%
      \expandafter\def\csname LTb\endcsname{\color{black}}%
      \expandafter\def\csname LTa\endcsname{\color{black}}%
      \expandafter\def\csname LT0\endcsname{\color{black}}%
      \expandafter\def\csname LT1\endcsname{\color{black}}%
      \expandafter\def\csname LT2\endcsname{\color{black}}%
      \expandafter\def\csname LT3\endcsname{\color{black}}%
      \expandafter\def\csname LT4\endcsname{\color{black}}%
      \expandafter\def\csname LT5\endcsname{\color{black}}%
      \expandafter\def\csname LT6\endcsname{\color{black}}%
      \expandafter\def\csname LT7\endcsname{\color{black}}%
      \expandafter\def\csname LT8\endcsname{\color{black}}%
    \fi
  \fi
  \setlength{\unitlength}{0.0500bp}%
  \begin{picture}(5668.00,4250.00)%
    \gplgaddtomacro\gplbacktext{%
      \csname LTb\endcsname%
      \put(897,205){\makebox(0,0){\strut{} 0}}%
      \put(1664,205){\makebox(0,0){\strut{} 0.25}}%
      \put(2431,205){\makebox(0,0){\strut{} 0.5}}%
      \put(3199,205){\makebox(0,0){\strut{} 0.75}}%
      \put(3966,205){\makebox(0,0){\strut{} 1}}%
      \put(2124,-125){\makebox(0,0){\strut{}$\varepsilon$}}%
    }%
    \gplgaddtomacro\gplfronttext{%
      \csname LTb\endcsname%
      \put(1587,3876){\makebox(0,0){\footnotesize Dashen's theorem}}%
    }%
    \gplgaddtomacro\gplbacktext{%
    }%
    \gplgaddtomacro\gplfronttext{%
      \csname LTb\endcsname%
      \put(4096,3635){\makebox(0,0)[l]{\strut{}\footnotesize{\citet{Duncan:1996cy}~(\citeyear{Duncan:1996cy}) (quenched QCD)}}}%
      \put(4096,3234){\makebox(0,0)[l]{\strut{}\footnotesize{RBC-UKQCD~\citep{Blum:2007gh}~(\citeyear{Blum:2007gh})}}}%
      \put(4096,2832){\makebox(0,0)[l]{\strut{}\footnotesize{\textit{BMWc~\citep{Portelli:2010tz}~(\citeyear{Portelli:2010tz})}}}}%
      \put(4096,2431){\makebox(0,0)[l]{\strut{}\footnotesize{RBC-UKQCD~\citep{Blum:2010ge}~(\citeyear{Blum:2010ge})}}}%
      \put(4096,2030){\makebox(0,0)[l]{\strut{}\footnotesize{\textit{BMWc~\citep{Portelli:2012vz}~(\citeyear{Portelli:2012vz})}}}}%
      \put(4096,1629){\makebox(0,0)[l]{\strut{}\footnotesize{RM123~\citep{deDivitiis:2013vc}~(\citeyear{deDivitiis:2013vc})}}}%
      \put(4096,1227){\makebox(0,0)[l]{\strut{}\footnotesize{\textit{MILC~\citep{Basak:2013ug}~(\citeyear{Basak:2013ug})}}}}%
      \put(4096,826){\makebox(0,0)[l]{\strut{}\footnotesize{BMWc~\citep{BMW:2013}~(\citeyear{BMW:2013})}}}%
    }%
    \gplbacktext
    \put(0,0){\includegraphics{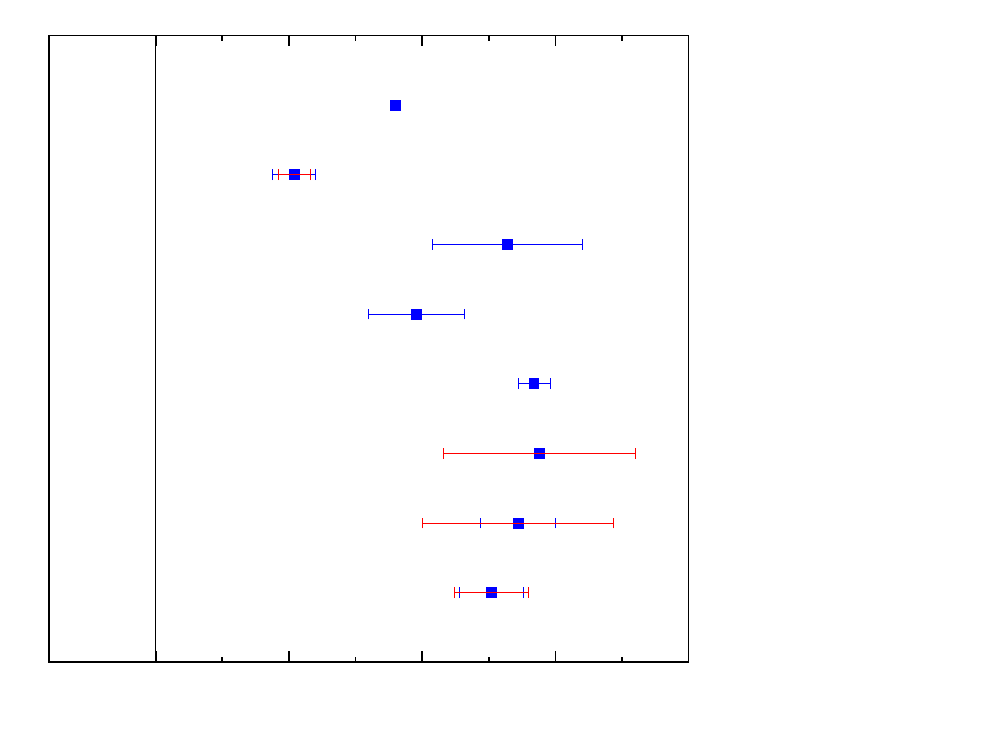}}%
    \gplfronttext
  \end{picture}%
\endgroup

%% file: figures/gp_comp_mq_lat_over.tex
\begingroup
  \makeatletter
  \providecommand\color[2][]{%
    \GenericError{(gnuplot) \space\space\space\@spaces}{%
      Package color not loaded in conjunction with
      terminal option `colourtext'%
    }{See the gnuplot documentation for explanation.%
    }{Either use 'blacktext' in gnuplot or load the package
      color.sty in LaTeX.}%
    \renewcommand\color[2][]{}%
  }%
  \providecommand\includegraphics[2][]{%
    \GenericError{(gnuplot) \space\space\space\@spaces}{%
      Package graphicx or graphics not loaded%
    }{See the gnuplot documentation for explanation.%
    }{The gnuplot epslatex terminal needs graphicx.sty or graphics.sty.}%
    \renewcommand\includegraphics[2][]{}%
  }%
  \providecommand\rotatebox[2]{#2}%
  \@ifundefined{ifGPcolor}{%
    \newif\ifGPcolor
    \GPcolortrue
  }{}%
  \@ifundefined{ifGPblacktext}{%
    \newif\ifGPblacktext
    \GPblacktexttrue
  }{}%
  \let\gplgaddtomacro\g@addto@macro
  \gdef\gplbacktext{}%
  \gdef\gplfronttext{}%
  \makeatother
  \ifGPblacktext
    \def\colorrgb#1{}%
    \def\colorgray#1{}%
  \else
    \ifGPcolor
      \def\colorrgb#1{\color[rgb]{#1}}%
      \def\colorgray#1{\color[gray]{#1}}%
      \expandafter\def\csname LTw\endcsname{\color{white}}%
      \expandafter\def\csname LTb\endcsname{\color{black}}%
      \expandafter\def\csname LTa\endcsname{\color{black}}%
      \expandafter\def\csname LT0\endcsname{\color[rgb]{1,0,0}}%
      \expandafter\def\csname LT1\endcsname{\color[rgb]{0,1,0}}%
      \expandafter\def\csname LT2\endcsname{\color[rgb]{0,0,1}}%
      \expandafter\def\csname LT3\endcsname{\color[rgb]{1,0,1}}%
      \expandafter\def\csname LT4\endcsname{\color[rgb]{0,1,1}}%
      \expandafter\def\csname LT5\endcsname{\color[rgb]{1,1,0}}%
      \expandafter\def\csname LT6\endcsname{\color[rgb]{0,0,0}}%
      \expandafter\def\csname LT7\endcsname{\color[rgb]{1,0.3,0}}%
      \expandafter\def\csname LT8\endcsname{\color[rgb]{0.5,0.5,0.5}}%
    \else
      \def\colorrgb#1{\color{black}}%
      \def\colorgray#1{\color[gray]{#1}}%
      \expandafter\def\csname LTw\endcsname{\color{white}}%
      \expandafter\def\csname LTb\endcsname{\color{black}}%
      \expandafter\def\csname LTa\endcsname{\color{black}}%
      \expandafter\def\csname LT0\endcsname{\color{black}}%
      \expandafter\def\csname LT1\endcsname{\color{black}}%
      \expandafter\def\csname LT2\endcsname{\color{black}}%
      \expandafter\def\csname LT3\endcsname{\color{black}}%
      \expandafter\def\csname LT4\endcsname{\color{black}}%
      \expandafter\def\csname LT5\endcsname{\color{black}}%
      \expandafter\def\csname LT6\endcsname{\color{black}}%
      \expandafter\def\csname LT7\endcsname{\color{black}}%
      \expandafter\def\csname LT8\endcsname{\color{black}}%
    \fi
  \fi
  \setlength{\unitlength}{0.0500bp}%
  \begin{picture}(5668.00,4250.00)%
    \gplgaddtomacro\gplbacktext{%
      \csname LTb\endcsname%
      \put(2124,-125){\makebox(0,0){\strut{}$m_u/m_d$}}%
    }%
    \gplgaddtomacro\gplfronttext{%
      \csname LTb\endcsname%
      \put(618,205){\makebox(0,0){\strut{} 0.3}}%
      \put(1287,205){\makebox(0,0){\strut{} 0.4}}%
      \put(1957,205){\makebox(0,0){\strut{} 0.5}}%
      \put(2627,205){\makebox(0,0){\strut{} 0.6}}%
      \put(3296,205){\makebox(0,0){\strut{} 0.7}}%
      \put(3966,205){\makebox(0,0){\strut{} 0.8}}%
      \put(1823,3830){\makebox(0,0){\footnotesize PDG 2012 \citep{PDG2012}}}%
    }%
    \gplgaddtomacro\gplbacktext{%
    }%
    \gplgaddtomacro\gplfronttext{%
      \csname LTb\endcsname%
      \put(4096,3520){\makebox(0,0)[l]{\strut{}\footnotesize{\citet{Duncan:1996cy}~(\citeyear{Duncan:1996cy}) (quenched QCD)}}}%
      \put(4096,3004){\makebox(0,0)[l]{\strut{}\footnotesize{RBC-UKQCD~\citep{Blum:2007gh}~(\citeyear{Blum:2007gh})}}}%
      \put(4096,2488){\makebox(0,0)[l]{\strut{}\footnotesize{RBC-UKQCD~\citep{Blum:2010ge}~(\citeyear{Blum:2010ge})}}}%
      \put(4096,1973){\makebox(0,0)[l]{\strut{}\footnotesize{PACS-CS~\citep{Aoki:2012uu}~(\citeyear{Aoki:2012uu})} (unquenched QED)}}%
      \put(4096,1457){\makebox(0,0)[l]{\strut{}\footnotesize{RM123~\citep{deDivitiis:2013vc}~(\citeyear{deDivitiis:2013vc})}}}%
      \put(4096,941){\makebox(0,0)[l]{\strut{}\footnotesize{BMWc~\citep{BMW:2013}~(\citeyear{BMW:2013})}}}%
    }%
    \gplbacktext
    \put(0,0){\includegraphics{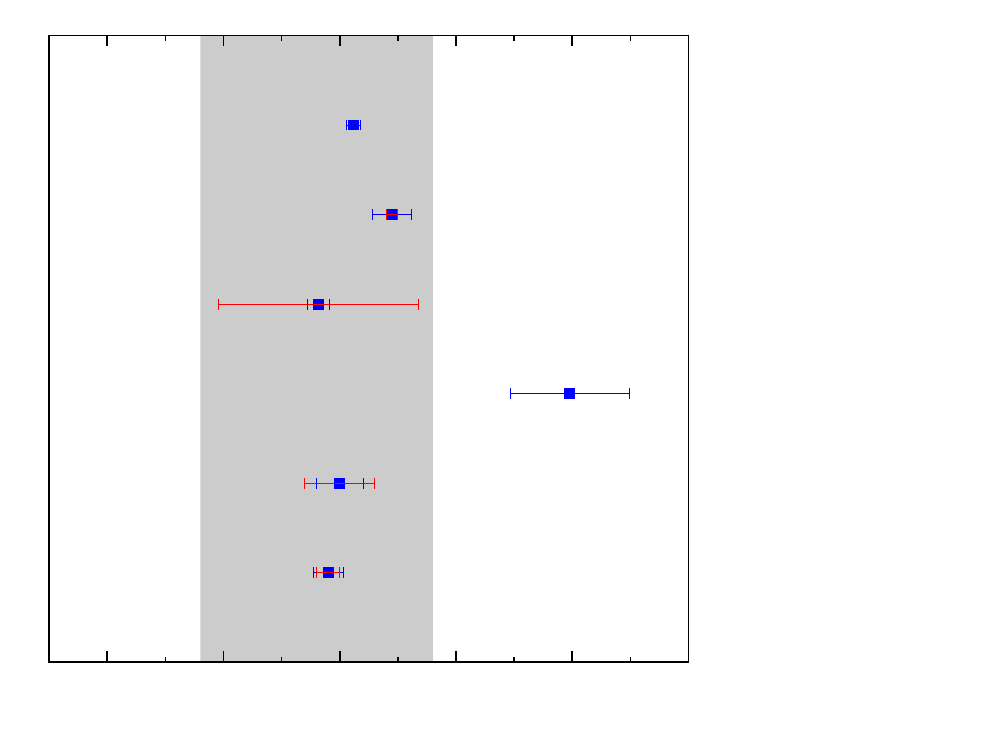}}%
    \gplfronttext
  \end{picture}%
\endgroup

%% file: figures/gp_comp_N_all_over.tex
\begingroup
  \makeatletter
  \providecommand\color[2][]{%
    \GenericError{(gnuplot) \space\space\space\@spaces}{%
      Package color not loaded in conjunction with
      terminal option `colourtext'%
    }{See the gnuplot documentation for explanation.%
    }{Either use 'blacktext' in gnuplot or load the package
      color.sty in LaTeX.}%
    \renewcommand\color[2][]{}%
  }%
  \providecommand\includegraphics[2][]{%
    \GenericError{(gnuplot) \space\space\space\@spaces}{%
      Package graphicx or graphics not loaded%
    }{See the gnuplot documentation for explanation.%
    }{The gnuplot epslatex terminal needs graphicx.sty or graphics.sty.}%
    \renewcommand\includegraphics[2][]{}%
  }%
  \providecommand\rotatebox[2]{#2}%
  \@ifundefined{ifGPcolor}{%
    \newif\ifGPcolor
    \GPcolortrue
  }{}%
  \@ifundefined{ifGPblacktext}{%
    \newif\ifGPblacktext
    \GPblacktexttrue
  }{}%
  \let\gplgaddtomacro\g@addto@macro
  \gdef\gplbacktext{}%
  \gdef\gplfronttext{}%
  \makeatother
  \ifGPblacktext
    \def\colorrgb#1{}%
    \def\colorgray#1{}%
  \else
    \ifGPcolor
      \def\colorrgb#1{\color[rgb]{#1}}%
      \def\colorgray#1{\color[gray]{#1}}%
      \expandafter\def\csname LTw\endcsname{\color{white}}%
      \expandafter\def\csname LTb\endcsname{\color{black}}%
      \expandafter\def\csname LTa\endcsname{\color{black}}%
      \expandafter\def\csname LT0\endcsname{\color[rgb]{1,0,0}}%
      \expandafter\def\csname LT1\endcsname{\color[rgb]{0,1,0}}%
      \expandafter\def\csname LT2\endcsname{\color[rgb]{0,0,1}}%
      \expandafter\def\csname LT3\endcsname{\color[rgb]{1,0,1}}%
      \expandafter\def\csname LT4\endcsname{\color[rgb]{0,1,1}}%
      \expandafter\def\csname LT5\endcsname{\color[rgb]{1,1,0}}%
      \expandafter\def\csname LT6\endcsname{\color[rgb]{0,0,0}}%
      \expandafter\def\csname LT7\endcsname{\color[rgb]{1,0.3,0}}%
      \expandafter\def\csname LT8\endcsname{\color[rgb]{0.5,0.5,0.5}}%
    \else
      \def\colorrgb#1{\color{black}}%
      \def\colorgray#1{\color[gray]{#1}}%
      \expandafter\def\csname LTw\endcsname{\color{white}}%
      \expandafter\def\csname LTb\endcsname{\color{black}}%
      \expandafter\def\csname LTa\endcsname{\color{black}}%
      \expandafter\def\csname LT0\endcsname{\color{black}}%
      \expandafter\def\csname LT1\endcsname{\color{black}}%
      \expandafter\def\csname LT2\endcsname{\color{black}}%
      \expandafter\def\csname LT3\endcsname{\color{black}}%
      \expandafter\def\csname LT4\endcsname{\color{black}}%
      \expandafter\def\csname LT5\endcsname{\color{black}}%
      \expandafter\def\csname LT6\endcsname{\color{black}}%
      \expandafter\def\csname LT7\endcsname{\color{black}}%
      \expandafter\def\csname LT8\endcsname{\color{black}}%
    \fi
  \fi
  \setlength{\unitlength}{0.0500bp}%
  \begin{picture}(7370.00,4250.00)%
    \gplgaddtomacro\gplbacktext{%
      \csname LTb\endcsname%
      \put(1565,-125){\makebox(0,0){\strut{}$\DQCDM_N~(\mathrm{MeV})$}}%
    }%
    \gplgaddtomacro\gplfronttext{%
      \csname LTb\endcsname%
      \put(767,205){\makebox(0,0){\strut{}-3.5}}%
      \put(1166,205){\makebox(0,0){\strut{}-3}}%
      \put(1566,205){\makebox(0,0){\strut{}-2.5}}%
      \put(1965,205){\makebox(0,0){\strut{}-2}}%
      \put(2364,205){\makebox(0,0){\strut{}-1.5}}%
    }%
    \gplgaddtomacro\gplbacktext{%
      \csname LTb\endcsname%
      \put(3960,-125){\makebox(0,0){\strut{}$\DQEDM_N~(\mathrm{MeV})$}}%
    }%
    \gplgaddtomacro\gplfronttext{%
      \csname LTb\endcsname%
      \put(3242,205){\makebox(0,0){\strut{} 0.5}}%
      \put(3721,205){\makebox(0,0){\strut{} 1}}%
      \put(4200,205){\makebox(0,0){\strut{} 1.5}}%
      \put(4679,205){\makebox(0,0){\strut{} 2}}%
    }%
    \gplgaddtomacro\gplbacktext{%
    }%
    \gplgaddtomacro\gplfronttext{%
      \csname LTb\endcsname%
      \put(5326,3635){\makebox(0,0)[l]{\strut{}\footnotesize{\citet{Gasser:1982ic}~(\citeyear{Gasser:1982ic})}}}%
      \put(5326,3234){\makebox(0,0)[l]{\strut{}\footnotesize{\citet{WalkerLoud:2012gq}~(\citeyear{WalkerLoud:2012gq})}}}%
      \put(5326,2832){\makebox(0,0)[l]{\strut{}\footnotesize{\citet{Shanahan:2013kg}~(\citeyear{Shanahan:2013kg})}}}%
      \csname LTb\endcsname%
      \put(5326,2431){\makebox(0,0)[l]{\strut{}\footnotesize{NPLQCD~\citep{Beane:2007eh}~(\citeyear{Beane:2007eh})}}}%
      \put(5326,2030){\makebox(0,0)[l]{\strut{}\footnotesize{RBC-UKQCD~\citep{Blum:2010ge}~(\citeyear{Blum:2010ge})}}}%
      \put(5326,1629){\makebox(0,0)[l]{\strut{}\footnotesize{QCDSF-UKQCD~\citep{Horsley:2012ue}~(\citeyear{Horsley:2012ue})}}}%
      \put(5326,1227){\makebox(0,0)[l]{\strut{}\footnotesize{RM123~\citep{deDivitiis:2013vc}~(\citeyear{deDivitiis:2013vc})}}}%
      \put(5326,826){\makebox(0,0)[l]{\strut{}\footnotesize{BMWc~\citep{Borsanyi:2013va}~(\citeyear{Borsanyi:2013va})}}}%
    }%
    \gplbacktext
    \put(0,0){\includegraphics{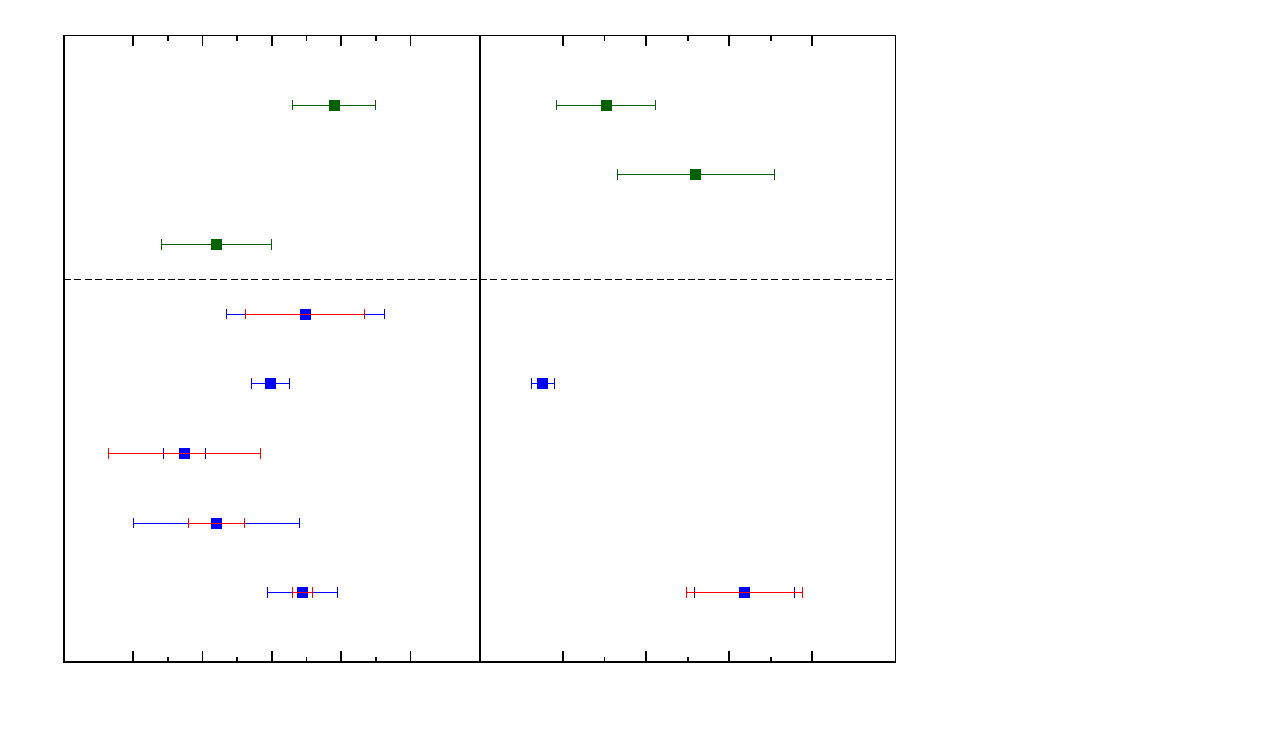}}%
    \gplfronttext
  \end{picture}%
\endgroup

%% file: figures/gp_final_split_over.tex
\begingroup
  \makeatletter
  \providecommand\color[2][]{%
    \GenericError{(gnuplot) \space\space\space\@spaces}{%
      Package color not loaded in conjunction with
      terminal option `colourtext'%
    }{See the gnuplot documentation for explanation.%
    }{Either use 'blacktext' in gnuplot or load the package
      color.sty in LaTeX.}%
    \renewcommand\color[2][]{}%
  }%
  \providecommand\includegraphics[2][]{%
    \GenericError{(gnuplot) \space\space\space\@spaces}{%
      Package graphicx or graphics not loaded%
    }{See the gnuplot documentation for explanation.%
    }{The gnuplot epslatex terminal needs graphicx.sty or graphics.sty.}%
    \renewcommand\includegraphics[2][]{}%
  }%
  \providecommand\rotatebox[2]{#2}%
  \@ifundefined{ifGPcolor}{%
    \newif\ifGPcolor
    \GPcolortrue
  }{}%
  \@ifundefined{ifGPblacktext}{%
    \newif\ifGPblacktext
    \GPblacktexttrue
  }{}%
  \let\gplgaddtomacro\g@addto@macro
  \gdef\gplbacktext{}%
  \gdef\gplfronttext{}%
  \makeatother
  \ifGPblacktext
    \def\colorrgb#1{}%
    \def\colorgray#1{}%
  \else
    \ifGPcolor
      \def\colorrgb#1{\color[rgb]{#1}}%
      \def\colorgray#1{\color[gray]{#1}}%
      \expandafter\def\csname LTw\endcsname{\color{white}}%
      \expandafter\def\csname LTb\endcsname{\color{black}}%
      \expandafter\def\csname LTa\endcsname{\color{black}}%
      \expandafter\def\csname LT0\endcsname{\color[rgb]{1,0,0}}%
      \expandafter\def\csname LT1\endcsname{\color[rgb]{0,1,0}}%
      \expandafter\def\csname LT2\endcsname{\color[rgb]{0,0,1}}%
      \expandafter\def\csname LT3\endcsname{\color[rgb]{1,0,1}}%
      \expandafter\def\csname LT4\endcsname{\color[rgb]{0,1,1}}%
      \expandafter\def\csname LT5\endcsname{\color[rgb]{1,1,0}}%
      \expandafter\def\csname LT6\endcsname{\color[rgb]{0,0,0}}%
      \expandafter\def\csname LT7\endcsname{\color[rgb]{1,0.3,0}}%
      \expandafter\def\csname LT8\endcsname{\color[rgb]{0.5,0.5,0.5}}%
    \else
      \def\colorrgb#1{\color{black}}%
      \def\colorgray#1{\color[gray]{#1}}%
      \expandafter\def\csname LTw\endcsname{\color{white}}%
      \expandafter\def\csname LTb\endcsname{\color{black}}%
      \expandafter\def\csname LTa\endcsname{\color{black}}%
      \expandafter\def\csname LT0\endcsname{\color{black}}%
      \expandafter\def\csname LT1\endcsname{\color{black}}%
      \expandafter\def\csname LT2\endcsname{\color{black}}%
      \expandafter\def\csname LT3\endcsname{\color{black}}%
      \expandafter\def\csname LT4\endcsname{\color{black}}%
      \expandafter\def\csname LT5\endcsname{\color{black}}%
      \expandafter\def\csname LT6\endcsname{\color{black}}%
      \expandafter\def\csname LT7\endcsname{\color{black}}%
      \expandafter\def\csname LT8\endcsname{\color{black}}%
    \fi
  \fi
  \setlength{\unitlength}{0.0500bp}%
  \begin{picture}(6236.00,3968.00)%
    \gplgaddtomacro\gplbacktext{%
      \csname LTb\endcsname%
      \put(682,576){\makebox(0,0)[r]{\strut{}$-9$}}%
      \put(682,848){\makebox(0,0)[r]{\strut{}$-8$}}%
      \put(682,1120){\makebox(0,0)[r]{\strut{}$-7$}}%
      \put(682,1392){\makebox(0,0)[r]{\strut{}$-6$}}%
      \put(682,1664){\makebox(0,0)[r]{\strut{}$-5$}}%
      \put(682,1936){\makebox(0,0)[r]{\strut{}$-4$}}%
      \put(682,2207){\makebox(0,0)[r]{\strut{}$-3$}}%
      \put(682,2479){\makebox(0,0)[r]{\strut{}$-2$}}%
      \put(682,2751){\makebox(0,0)[r]{\strut{}$-1$}}%
      \put(682,3023){\makebox(0,0)[r]{\strut{}$0$}}%
      \put(682,3295){\makebox(0,0)[r]{\strut{}$1$}}%
      \put(682,3567){\makebox(0,0)[r]{\strut{}$2$}}%
      \put(1652,220){\makebox(0,0){\strut{}$\Delta M_{N}$}}%
      \put(3327,220){\makebox(0,0){\strut{}$\Delta M_{\Sigma}$}}%
      \put(5002,220){\makebox(0,0){\strut{}$\Delta M_{\Xi}$}}%
      \put(176,2071){\rotatebox{-270}{\makebox(0,0){\strut{}$(\mathrm{MeV})$}}}%
      \put(2573,3567){\makebox(0,0)[l]{\strut{}\footnotesize{Budapest-Marseille-Wuppertal collaboration}}}%
    }%
    \gplgaddtomacro\gplfronttext{%
      \csname LTb\endcsname%
      \put(1405,1273){\makebox(0,0)[l]{\strut{}total}}%
      \csname LTb\endcsname%
      \put(1405,1053){\makebox(0,0)[l]{\strut{}QCD}}%
      \csname LTb\endcsname%
      \put(1405,833){\makebox(0,0)[l]{\strut{}QED}}%
      \csname LTb\endcsname%
      \put(1405,613){\makebox(0,0)[l]{\strut{}exp.}}%
    }%
    \gplbacktext
    \put(0,0){\includegraphics{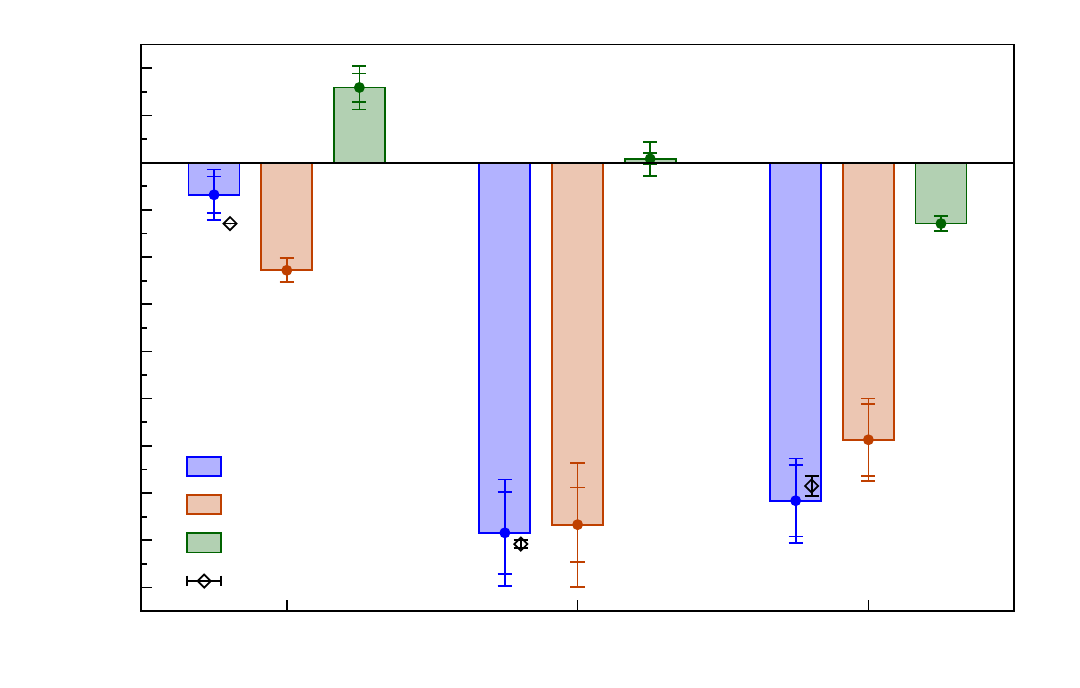}}%
    \gplfronttext
  \end{picture}%
\endgroup